\documentclass[aps,prl,floatfix,superscriptaddress,twocolumn,footinbib]{revtex4-2}

\usepackage{amssymb}
\usepackage{amsmath}
\usepackage{amsfonts}
\usepackage{appendix}
\usepackage{bm}
\usepackage{braket}
\usepackage{graphicx}
\usepackage{epsfig}
\usepackage{epstopdf}
\usepackage{balance}
\usepackage[dvipsnames]{xcolor}
\usepackage{calc}
\usepackage{natbib}
\usepackage[colorlinks,
            linkcolor=blue,
            anchorcolor=blue,
            citecolor=blue,
            urlcolor=blue]{hyperref}
\usepackage{lipsum}

\begin{document}
\title{Theory of Localized States in Quasiperiodic Lattices
}

\author{Jin-Rong Chen}
	\affiliation{School of Physics and Wuhan National High Magnetic Field Center, Huazhong University of Science and Technology, Wuhan, Hubei 430074, China}
  \author{Xin-Yu Guo}
	\affiliation{School of Physics and Wuhan National High Magnetic Field Center, Huazhong University of Science and Technology, Wuhan, Hubei 430074, China}
 \author{Shi-Ping Ding}
	\affiliation{School of Physics and Wuhan National High Magnetic Field Center, Huazhong University of Science and Technology, Wuhan, Hubei 430074, China}
 
  \author{Tian-Le Wu}
	\affiliation{School of Physics and Wuhan National High Magnetic Field Center, Huazhong University of Science and Technology, Wuhan, Hubei 430074, China}
 
  \author{Miao Liang}
  \affiliation{School of Physics and Wuhan National High Magnetic Field Center, Huazhong University of Science and Technology, Wuhan, Hubei 430074, China}
      \author{Jin-Hua Gao}
 \email{jinhua@hust.edu.cn}
	\affiliation{School of Physics and Wuhan National High Magnetic Field Center, Huazhong University of Science and Technology, Wuhan, Hubei 430074, China}
 \author{X. C. Xie}
 \affiliation{Interdisciplinary Center for Theoretical Physics and Information Sciences (ICTPIS), Fudan University
, Shanghai 200433, China}
\affiliation{International Center for Quantum Materials, School of Physics, Peking University, Beijing 100871, China}
 \affiliation{Hefei National Laboratory, Hefei 230088, China}
\begin{abstract}
The physics of localized states in quasiperiodic lattices has been extensively studied for decades, but still lacks an comprehensive theoretical framework.
 Recently, we developed a incommensurate energy band (IEB) theory, which extends the concept of energy bands to quasiperiodic systems lacking translational symmetry, thereby achieving a breakthrough in elucidating extended states.
 Here, we demonstrate that, due to the inherent duality between momentum and real space, the IEB theory also offers a comprehensive framework for elucidating localized states. Specifically, via a so-called \emph{spiral (module) mapping}, the energy spectrum of localized states can be represented as a function defined on a compact circular manifold—akin to the Brillouin zone—whose form resembles conventional energy bands. These \emph{localized state energy bands} (LSEBs) fully characterize all the properties of the localized states. Moreover, we show that quasiperiodic systems with mobility edges exhibit a unique hybrid band structure:  the IEB for extended states (momentum space) and LSEB for localized states (real space), separated by mobility edges. Our theory thus establishes a comprehensive framework for analyzing the localized states in quasiperiodic lattices.

\end{abstract}
\maketitle

\emph{Introduction.}---
Electron localization in solids is a fundamental phenomenon in condensed matter physics. Unlike delocalized Bloch electrons, localized quantum states arise due to quasiperiodic potentials or disorder (Anderson localization)\cite{Anderson,Thouless,Mott_1987,RevModPhys.80.1355}. A paradigmatic example is the Aubry-André-Harper(AAH) model\cite{aubry_1980,Harper_1955}, which describes a one-dimensional  atomic chain  (lattice constant $a_0$)  subjected to a quasiperiodic potential with periodicity $b_0$, where the ratio $\alpha \equiv a_0/b_0$ is irrational. In this model, electron localization occurs when the potential strength exceeds a critical value determined by the Aubry-André self-duality\cite{wilkinson1984critical,A.P.Siebesma_1987,PhysRevA.36.5349,almost_mathieu_operator,PhysRevLett.51.1198}. Intriguingly, if the self-duality condition is broken—for instance, in generalized AAH (GAAH) models with modified quasiperiodic potentials—a mobility edge emerges in the energy spectrum, marking a sharp boundary between localized and extended states\cite{PhysRevLett.114.146601,PhysRevLett.125.196604}.


Despite over half a century of intensive research on quasiperiodic lattices\cite{PhysRevLett.114.146601,Nature07,Roati2008anderson,Nature_Billy,PhysRevLett.104.070601,PhysRevLett.125.196604,PhysRevLett.123.025301,PhysRevLett.122.237601,goblot2020emergence,PhysRevLett.126.106803,PhysRevB.109.014210,PhysRevLett.132.236301,PhysRevLett.48.1043,PhysRevLett.123.070405,PhysRevB.91.014108,PhysRevLett.110.176403,PhysRevB.93.104504,PhysRevB.100.125157,PhysRevB.104.014202,PhysRevB.96.054202,PhysRevB.106.024204,PhysRevB.105.014207,PhysRevB.99.054211,PhysRevLett.131.176401,PhysRevLett.110.180403,PhysRevB.87.134202,PhysRevLett.113.045304,PhysRevB.28.4272,PhysRevLett.61.2144,PhysRevB.41.5544,PhysRevB.96.085119,PhysRevLett.98.130404,PhysRevLett.109.106402,PhysRevLett.120.160404,PhysRevLett.61.2141, PhysRevB.103.L060201,PhysRevB.106.054204,PhysRevB.107.085111,PhysRevB.107.075128,PhysRevLett.60.1334,PhysRevB.98.104201,PhysRevB.108.174206,PhysRevLett.101.076803,10.1007/s11511-015-0128-7,PhysRevLett.131.186303,PhysRevB.108.L100201,PhysRevLett.115.186601,PhysRevB.103.174205,PhysRevLett.103.013901,Modugno_2009,PhysRevLett.109.106402,science,PhysRevLett.120.160404,PhysRevB.14.2239,PhysRevB.42.8282,hakan1997,avila2005,avila2015,zhouqi2015,PhysRevB.110.L060201,PhysRevB.94.125408,PhysRevB.104.224204,AAH_cold_atom1,PhysRevLett.108.220401,PhysRevB.101.174205,PhysRevB.111.214205}, current theoretical approaches to their energy spectrum calculation are fundamentally limited, compared to the maturity achieved in periodic lattice systems. In most scenarios, we can only determine the energy spectrum of quasiperiodic systems—whether extended or localized states—by numerically diagonalizing a real-space lattice made as large as practically achievable\cite{PhysRevLett.114.146601,PhysRevB.109.014210,PhysRevLett.132.236301,PhysRevLett.108.220401,goblot2020emergence,rl1f-ptzq,PhysRevLett.125.196604,PhysRevLett.123.025301,PhysRevLett.48.1043,PhysRevLett.61.2144,PhysRevLett.123.070405,PhysRevLett.104.070601,PhysRevLett.126.106803,PhysRevLett.122.237601,PhysRevB.91.014108,PhysRevResearch.2.033052,PhysRevB.41.5544,PhysRevA.103.043325,PhysRevB.103.134208,PhysRevB.101.174205,PhysRevB.104.024201,PhysRevB.111.214205,PhysRevB.110.134203,PhysRevA.110.012222,PhysRevLett.126.106803,PhysRevLett.110.176403,PhysRevB.93.104504,PhysRevB.100.125157,PhysRevB.104.014202,PhysRevB.96.054202,PhysRevB.106.024204,PhysRevB.105.014207,PhysRevB.99.054211,PhysRevLett.131.176401,PhysRevB.96.085119,PhysRevB.83.075105}.  The main reason is that, for the extended states, due to the lack of overall translational symmetry, the Bloch theory as well as the concept of energy band is inapplicable for the quasiperiodic systems. Equally formidable is the challenge that a comprehensive theoretical framework of localized states remains critically underdeveloped.

Recently, the energy spectrum theory of quasiperiodic lattices has achieved an essential breakthrough, especially on the extended states\cite{he2024energy,guo}. In our recent work, we introduce the concept of incommensurate energy band (IEB), a generalization of the conventional energy band into quasiperiodic (or incommensurate) systems without the need for translational symmetry. It is proved that the spectrum of extended states in quasiperiodic lattices (e.g., the AAH model) can be conveniently described as IEB, which not only has nearly the same form and calculation procedure as conventional energy bands but also recovers conventional energy bands when $\alpha$ is rational\cite{guo}.

In this work, we address the remaining challenge, namely, developing a comprehensive theoretical framework for the localized states in quasiperiodic lattices. 
The key point lies in the duality relationship between momentum space and real space\cite{aubry_1980,PhysRevLett.48.1043,PhysRevB.43.13468,PhysRevLett.51.1198,Sun_2015}. By considering this duality and taking Bloch functions as basis, the AAH model can also be equivalently formulated as a quasiperiodic lattice in momentum space. Consequently, the IEB theory can be applied to this momentum-space quasiperiodic lattice, thereby constructing an IEB framework specifically tailored for localized states. An intriguing aspect here is that, within such localized state energy band (LSEB) theory, the real-space lattice can be reduced via a simple spiral (or modulo) mapping to a circle parameterized by reduced coordinates, analogous to how crystal momenta are reduced to the Brillouin zone (BZ). Remarkably, this circle precisely serves a role analogous to the Brillouin zone. We then illustrate that the LSEB theory accurately captures all properties of localized states in quasiperiodic lattices, e.g.,  the density of states (DOS) and energy gaps.
Finally, we propose that for quasiperiodic lattices with mobility edges (e.g., the GAAH model\cite{PhysRevLett.114.146601}), their energy spectrum exhibits a unique hybrid band structure: bounded by the mobility edges, extended states can be described by the IEB theory in momentum space, while localized states are captured by the LSEB framework in real space.

This work, together with the proposed IEB theory\cite{he2024energy,guo}, establishes a comprehensive theoretical framework for the energy spectrum of quasiperiodic lattices.


\begin{figure}[t!]
    \centering
    \includegraphics[scale=0.32]{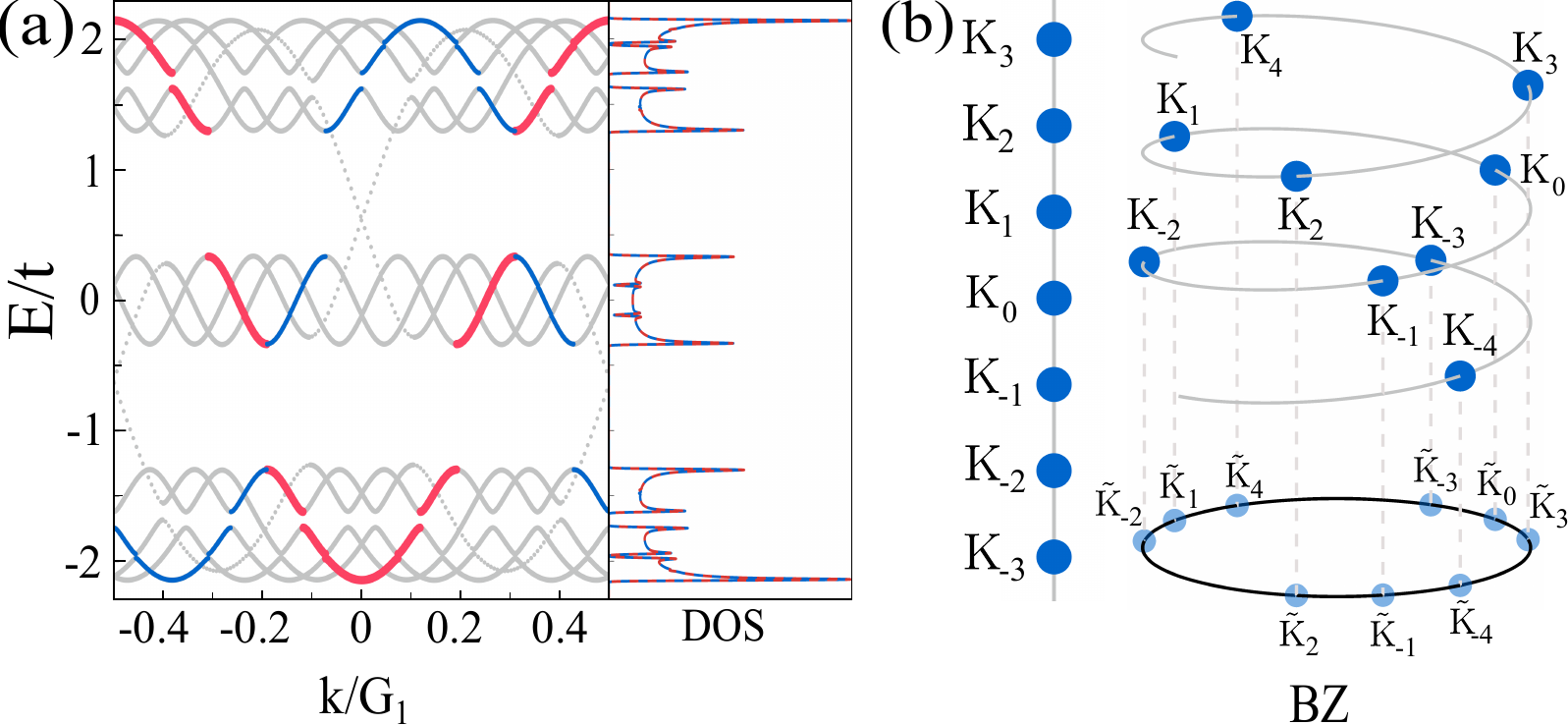}
    \caption{
    (a)~IEB (red lines) of AAH model with $\alpha = \frac{\sqrt{5}-1}{2}$, $\vartheta = 0$, $t=-1$, $V_0=t$ and $n_c=6$. Blue and gray lines show the replica bands. Right panel give the calculated DOS got by IEB (red) and diagonalizing large lattice (blue). 
    (b)~Schematic of the spiral (module) mapping. Left panel shows the coupled chain of Bloch states $|K_m\rangle$ with lattice constant $G_2$, Right panel illustrates the wrapping process of the coupled chain around the BZ circle, where $\widetilde{K}_m=K_m \mod G_1$.}
    \label{fig1}
\end{figure}

\emph{IEB theory.}---
Due to the duality between real  and momentum spaces, the theory for localized states exhibits formal similarity to the recently proposed IEB  theory for extended states. Therefore, let us start with a short introduction to the concept of IEB with the pedagogical AAH model. 

The  Hamiltonian of AAH model is 
\begin{equation}{\label{AAH_real_space_Lattice}}
    H_{\mathrm{AAH}}=t \sum_{j}( c^{\dagger}_{j}c_{j+1}+\mathrm{H.c.})+\sum_jV_jn_j 
\end{equation}
where $c_j$ is the electron annihilation operation on site $j$ and the quasiperiodic potential $V_j=V_0 \cos(\mathbf{G}_2 \cdot \mathbf{r}_j + \theta )$.  $\mathbf{G}_2=2\pi/b_0$ ($\mathbf{G}_1=2\pi/a_0$) is the reciprocal lattice vector of quasiperidic potential (atomic chain).  $\theta$ is a potential parameter. 

The IEB is defined in momentum space. Thus, we switch to the Bloch basis $|k\rangle = \frac{1}{\sqrt{N}} \sum_j \exp(ik\cdot R_j) |j\rangle$. Here, $|j\rangle$ and $R_j$ are the atomic orbital and position of $j$-th atom, respectively.
Then, the Hamiltonian~\eqref{AAH_real_space_Lattice} becomes 
\begin{equation}{\label{AAH_k_space_BZ}}
    H_{\mathrm{AAH}}=\frac{V_0}{2} \sum_{k\in \left[0,G_1 \right) }( c^{\dagger}_{k}c_{k+G_2}+\mathrm{H.c.})+\sum_{k \in \left[0,G_1 \right) } T_kn_k 
\end{equation}
where $c_k$ is the electron annihilation operator of the Bloch state $|k\rangle$, $T_k=2t\cos(k\cdot a_0)$ is the kinetic energy of $|k\rangle$.
Eq.~\eqref{AAH_k_space_BZ} shows that the quasiperiodic potential $V_j$ couples only the Bloch waves in the set $Q_k = \{ |K_m\rangle : K_m \equiv k + m G_2, m \in \mathbb{Z} \}$.
Note that, although the crystal momentum  $k$  is no longer a good quantum number here, we can still define energy bands for incommensurate systems. This is because constituting a well-defined energy band only requires using $k$ to label all eigenstates — without $k$ needing to be a good quantum number.

Let us summarize the basis steps  to get a IEB for the AAH model:

\begin{enumerate}
  \item For a given $k$, the Hamiltonian of AAH model can be expressed as an infinite-dimensional matrix in the basis $|K_m\rangle$ with $m \in \mathbb{Z}$\cite{supplemental1}. Following truncation with a cutoff $|m|\leq m_c$ and diagonalization to obtain all eigenstates, the fundamental question for constructing the IEB reduces to: which eigenstate properly corresponds to the input $k$?
  
\item In fact, for the Bloch basis $|K_m\rangle$ associated with the Hamiltonian matrix, only $|k\rangle$ with $m=0$ corresponds to the real physical system, while all other basis states  represent replica bands introduced to handle the incommensurate coupling. This becomes particularly clear when we turn off the coupling ($V_0=0$): the AAH model reverts to a one-dimensional atomic chain where $|k\rangle$ becomes its exact eigenstate, whereas all other $|K_m\rangle$  with $m \neq 0$ constitute replica bands yielding redundant eigenstates. This scenario is analogous to the Nambu representation formalism in BCS superconductivity theory\cite{Altland_Simons_2010} – both cases employ artificially introduced redundant degrees of freedom to handle coupling between electrons.

\item Note that we focus on extended states here, which are localized in momentum space. Consequently, when the coupling $V_0$ is turned on, although the true eigenstate develop broadening in k-space, its wavefunction remains predominantly localized near the basis state $|k\rangle$. More generally, for extended states, each basis $|K_m\rangle$ corresponds to an eigenstate whose wavefunction is primarily localized around $|K_m\rangle$ in momentum space. Thus, by selecting the eigenstate whose the maximum weight component resides near $|k\rangle$, we can construct the desired IEB for the AAH model, i.e., the eigenvalue as a function of $k$. 
\end{enumerate}

The IEB is in fact a natural generalization of energy band concept, which  will revert to the conventional energy band in the commensurate cases\cite{guo}. The calculated IEB for the AAH model is plotted in Fig.~\ref{fig1} (a), where red solid lines are the IEB corresponding to $|k\rangle$ and blue (gray) lines represent one (other) replica bands. The calculated DOS with IEB (red) is also plotted, which exactly coincide to that got by diagonalizing a large enough lattice (blue).   

\begin{figure*}[t!]
    \includegraphics[scale=0.37]{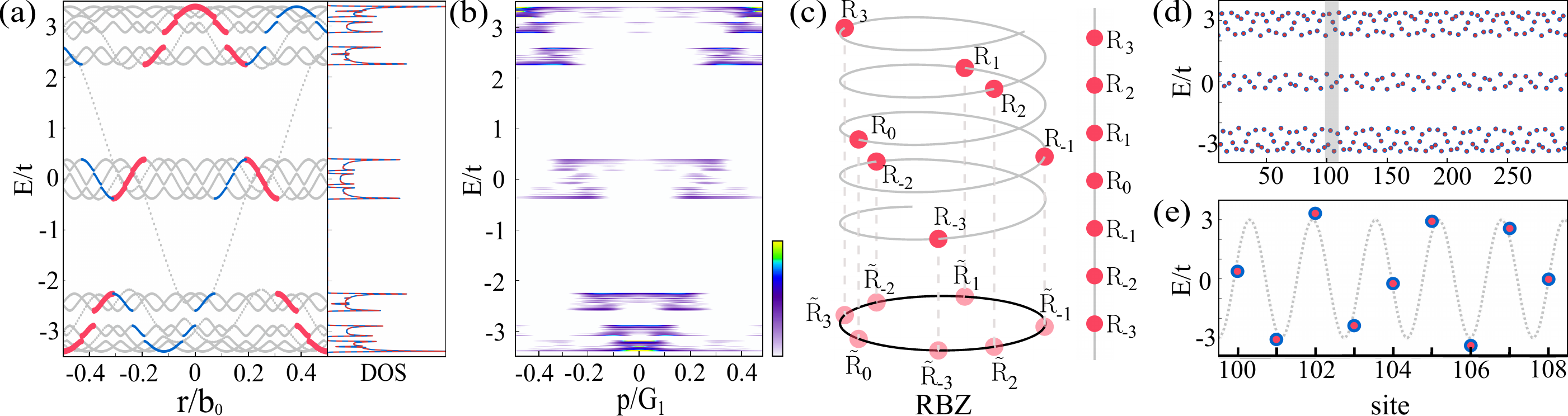}
    \caption{  
    (a)~Localized states energy band(LSEB)(red lines) of AAH model with $n_c=9$, the pareameters of (a-e) are $\alpha = (\sqrt{5}-1)/{2}$, $\vartheta = 0$, $t=-1$, $V_0=3t$. Blue and gray lines show the replica bands. Right panel give the calculated DOS denoted by LSEB(red) and diagonalizing large lattice(blue).
    (b)~is the calculated ARPES spectra $\sqrt{I}$, with $n_c=150$\cite{supplemental1}.
    (c)~Schematic of the spiral (module) mapping in real space. Left panel illustrates the wrapping process of the real lattice around the Real-space BZ(RBZ) circle, Right panel shows the coupled chain of atom orbital functions $\ket{R_j}$ with lattice constant $a_0$.
    (d)~The real-space spectrum comparison of two methods, the LSEB(red) with $n_c=12$ and the numerical diagonalization result(blue) with $L=500$. Specifically, The red energies are shifted from the LSEB in RBZ via the Spiral Mapping relation, while the blue energies are located in specific site according to their eigenstates' distribution.
    (e) Magnified view of the gray rectangle in (d). The dotted gray curved lines outline the incommensurate potential $V_0$. }
    \label{fig2}
\end{figure*}

\emph{Spiral mapping in momentum space}.---The other pivotal concept constituting the localized states theory is the  spiral (or module) mapping  relation, existing in both momentum and real space. 
For pedagogical clarity, we first explain the concept of spiral mapping in momentum space.

As shown in Eq.~\eqref{AAH_k_space_BZ}, the quasiperiodic potential $V_j$ couples infinitely many Bloch states $|K_m\rangle$, forming a 1D chain in momentum space indexed by $m$. Since Bloch waves require crystal momenta confined to the BZ $[0,G_1)$, each $K_m$ is projected to a reduced crystal momentum $\widetilde{K}_m \equiv K_m \mod G_1$ within the BZ. Crucially, when $G_2/G_1$ is irrational, the set $\{ \widetilde{K}_m\}$  densely and non-repetitively covers the entire BZ---a hallmark of incommensurate systems (ergodic feature).
This dense coverage enables a key numerical simplification: The complete set $\{\widetilde{K}_m\}$ can approximate the summation over the BZ. Consequently, we adopt the following \emph{ansatz}: Bloch states on the coupled chain ($\ket{K_m}$) provide a approximate representation for all states in the BZ.
Physically, this implies that the quasiperiodic potential couples every state across the BZ. Under this ansatz, each Bloch state admits dual equivalent labels: (1) reduced crystal momentum $k \in [0,G_1)$; 
(2) site index $m$  on the coupled chain with $k=\widetilde{K}_m$. 
Note that, in order to include $\Gamma$ point, we then always assume $K_m=mG_2$.

Fig.~\ref{fig1}(b) graphically demonstrates how the Bloch states on the coupled chain densely cover the BZ.
The circle with circumference $G_1$ denotes the BZ. When we wrap the coupled chain (lattice constant $G_2$) around this circle, the lattice sites $K_m$ are precisely mapped to the positions $\widetilde{K}_m$ within the BZ. Finally, all lattice points $\widetilde{K}_m$  densely cover the entire circle.
Representing this wrapping as a spiral configuration clarifies such module mapping relation, which we designate as \emph{spiral mapping}---a correspondence between the Bloch state indices $k \in \textrm{BZ}$ and $m \in \mathbb{Z}$.  This is just the concept of spiral mapping in momentum space.


\emph{ Aubry-André self-duality}.---The spiral mapping provides a very intuitive physical interpretation of the Aubry-André self-duality transformation\cite{aubry_1980,PhysRevLett.51.1198,Sun_2015}. 

With spiral mapping,  we can reformulate the Eq.~\eqref{AAH_k_space_BZ} via using the site index $m$ of coupled chain as the labels of Bloch states:
\begin{equation}{\label{AAH_k_space_Lattice}}
    H_{\mathrm{AAH}}=\frac{V_0}{2} \sum_{m }( c^{\dagger}_{m}c_{m+1}+\mathrm{H.c.})+\sum_{m} T_mn_m,
\end{equation}
where $c_m \equiv c_{K_m}$ and $T_m \equiv 2t \textrm{cos} (K_m \cdot a_0)$.
Importantly, Eq.\eqref{AAH_k_space_Lattice} is exactly the  standard self-duality form of the AAH model, which demonstrates that the AAH model admits completely analogous quasiperiodic 1D lattice structures in both momentum space and real space, differing only in the interchange of parameters $t$ and $\frac{V_0}{2}$. 
 It thus reveals the nature of the self-dual transformation in the AAH model: In essence, it is nothing but a change of basis from Wannier states to Bloch states, which merely employs the coupled chain index to label the Bloch states.

\emph{Localized state theory}.---Let's now turn to the theory of localized states. In the AAH model, when $V_0>2t$, eigenstates undergo a transition from extended to localized states\cite{PhysRevLett.48.1043,PhysRevLett.51.1198,PhysRevLett.104.070601,Sun_2015}. Once the electron localization emerges, the IEB approach breaks down\cite{supplemental1}. The reason is that, since these localized states are extended in momentum space,  we can no longer select eigenstates based on their momentum-space wavefunction distribution.

Fortunately, even in the localized regime, Equation \eqref{AAH_k_space_Lattice} still holds, which means that  the AAH model is equivalent to a 1D quasiperiodic lattice in momentum space. Because  electrons now behave as extended states in  momentum space, we can construct an analogous IEB theory on the momentum lattice to characterize the spectrum of the localized states, just exchanging  the roles of real and momentum space. This is the key idea of our localized state theory.

To develop the  localized state theory, a crucial point is recognizing that in real space, each Wannier state of a quasiperiodic lattice also possesses two distinct labeling schemes---analogous to Bloch states---with a real-space spiral mapping connecting these indices. 
In Fig.~\ref{fig2} (c), we illustrate this real-space spiral mapping. The position of each lattice site is given by $R_j = j a_0$, where $j$ serves as the site index for Wannier functions $|j\rangle$. When wrapping the 1D atomic chain (lattice constant $a_0$) onto a circle of circumference $b_0$, $R_j$ maps to a reduced coordinate
$\widetilde{R}_j \equiv R_j \mod b_0.$
Similar to the momentum-space case, the set $\{ \widetilde{R}_j \}$ densely covers the circle, when $\alpha$ is irrational. Consequently, owing to this spiral mapping relationship, the reduced coordinate $r \equiv \widetilde{R}_j \in [0,b_0)$ can also label Wannier state $|r\rangle=|j\rangle$. Furthermore, for localized states, this circle---namely, the interval $[0,b_0)$ where the reduced coordinate live---serves as an effective Brillouin zone in real space (RBZ). As will be shown later, the energy spectrum of localized states is defined on this RBZ.

Then, we can rewrite Eq.~\eqref{AAH_real_space_Lattice} with the reduced coordinate index,
\begin{equation}{\label{AAH_real_space_BZ}}
    H_{\mathrm{AAH}}= \sum_{r \in \textrm{RBZ}}t( c^{\dagger}_{r}c_{r+a_0}+\mathrm{H.c.})+\sum_{r \in \textrm{RBZ}} V_r n_r
\end{equation}
which  exactly has the same  form as Eq.~\eqref{AAH_k_space_BZ}. Eq.~\eqref{AAH_real_space_BZ} can also be obtained via basis transformation. Considering the relation, $|j\rangle=\frac{1}{\sqrt{N}}\sum_{k \in \textrm{BZ}} e^{-ik\cdot R_j} |k\rangle$, if we label the Wannier state $|j\rangle$ by the reduced coordinate  $r=\widetilde{R}_j$,  and denote the Bloch wave $|k\rangle$  by the coupled chain index $K_m$ ($k=\widetilde{K}_m$),  the transformation relation becomes $|r\rangle = \frac{1}{\sqrt{N}} \sum_{m} e^{-i K_m \cdot r} |K_m\rangle$. Here, the relation $e^{-ik\cdot R_j}=e^{-iK_m\cdot r}$ is used\cite{supplemental1}. Using this transformation, the momentum lattice in Eq.~\eqref{AAH_k_space_Lattice} can be directly changed into Eq~\eqref{AAH_real_space_BZ}.

\begin{figure*}[t!]
    \includegraphics[scale=0.54]{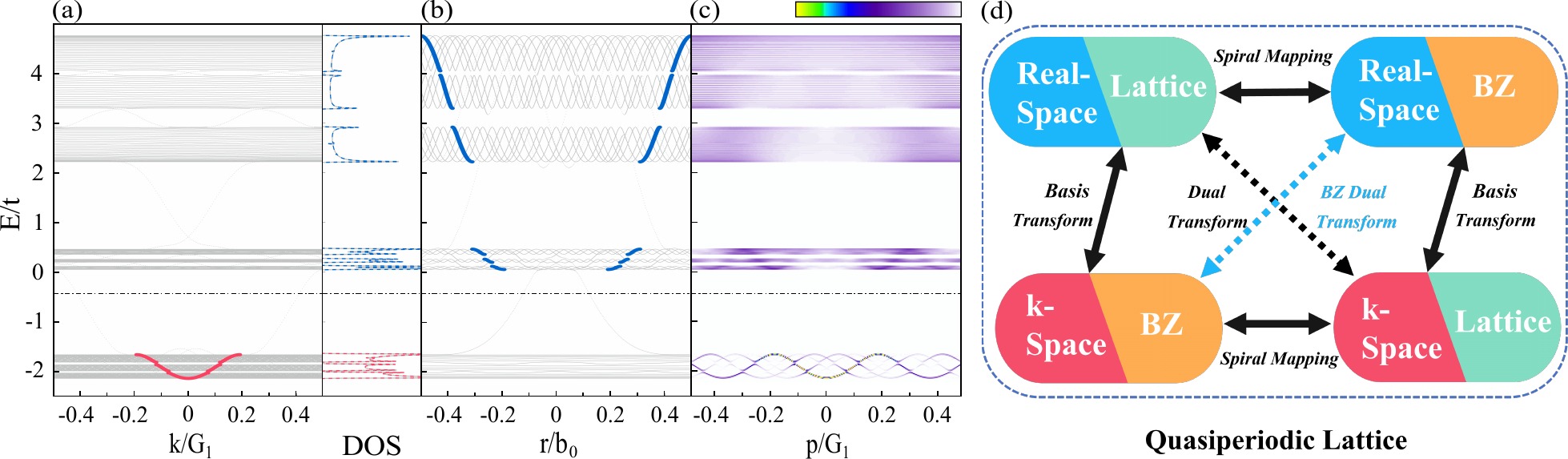}
    \caption{
    (a-b) is the calculated hybrid energy spectra of GAAH model , with $\alpha=(\sqrt{5}-1)/{2}$, $\vartheta=0$, $\beta=-0.5$, $\lambda=-1.1t$. In (a)((b)), the red(blue) lines are the IEB(LSEB) with $n_c=100$($n_c=20$), $n_e=5$($n_e=2$). The gray lines below(above) the dashed line in (a)((b)) are the replica bands of IEB(LSEB), while the other grey lines show the multiple energy spectrum. The panel between (a) and (b) denotes the DOS comparison between hybrid spectrum and numerical diagnolization(gray), the red(blue) DOS represents IEB(LSEB), respectively. 
    (c) is the calculated ARPES spectra $\sqrt{I}$ for (a) with $n_c=100$, $n_e=5$. The black dotted lines across (a)-(c) refers to the Mobility Edge $E_{ME}=-0.4$.
    (d) is the overall framework for localized states theory in quasiperiodic lattices. It introduces four equivalent representations of the quasiperiodic lattice Hamiltonian and the relationships connecting them. The dashed lines refer to the dual transform in AAH model\cite{supplemental1}.
    }
    \label{fig3}
\end{figure*}
Now, based on Eq.~\eqref{AAH_real_space_BZ}, the energy spectrum $E(r)$ of localized states in the AAH model can be calculated via exactly the same procedure as in IEB theory\cite{supplemental1}. Fig.~\ref{fig2} (a)  presents the numerical results of $E(r)$ on RBZ  (red lines), exhibiting a band structure exactly resembling IEB. 
Note that, in the AAH model, each localized state predominantly localizes near a single lattice site, establishing a one-to-one correspondence between localized states and lattice sites. This allows us to uniquely label localized states using lattice site indices. Therefore, an intuitive picture of $E(r)$ is: the eigenenergy of the localized state around the site $j$ with $r=\widetilde{R}_j$. Given the duality relation, for localized states, $E(r)$ is analogous to the concept of traditional energy bands, which  can fully characterize all properties of the localized states of the quasiperiodic systems.  So, we name $E(r)$ \emph{the localized state energy band} (LSEB), being the central result  of this work.  

$E(r)$ gives the correct DOS. In Fig.~\ref{fig2} (a),  we calculate the DOS of the localized states  with $E(r)$ (red lines)\cite{supplemental1}, 
\begin{equation}{\label{real_space_BZ_DOS}}
    \rho(\epsilon)=\lim_{L \rightarrow \infty} \dfrac{1}{L} \sum_{j=0}^{L-1} \delta(\epsilon-E(R_j))=\frac{1}{b_0} \int_{0}^{b_0} dr \delta(\epsilon-E(r)).
\end{equation}
Meanwhile, DOS can also be calculated via a direct diagonalization of a large enough AAH lattice (blue lines), which is also plotted in Fig.~\ref{fig2} (a). We see that the DOS got from two distinct methods perfectly agree with each other. Note that a great advantage of our method is that it can achieve high-accuracy results without the need to diagonalize a large matrix.

$E(r)$ exhibits unique ARPES spectrum of the localized states in quasiperiodic lattices.  In Fig.~\ref{fig2} (b), we calculate ARPES spectrum of the $E(r)$ of the AAH model, which markedly differ from those of conventional Bloch waves. The reason is that  the extension of localized state wave functions in momentum space leads to a highly dispersive shape of the ARPES spectrum.

$E(r)$ accurately determines the spatial distribution of localized states. 
 In Fig.~\ref{fig2} (d), we first plot the eigenenergy of the localized states as a function the lattice site index (red dots), using the relation $E(r=\widetilde{R}_j)$ for $j$-th site. Then, the eigenenergy $E(R_j)$ is calculated via directly diagonalizing a larger enough AAH lattice, see the blue dots in Fig.~\ref{fig2} (d).
 The exact consistency between the blue and red dots demonstrates that the LSEB  does rigorously characterize  spatial configurations of the localized states in quasiperiodic lattice.

Finally, we emphasize that $E(r)$ in fact provide a comprehensive framework to analyzing the properties of the localized states in quasiperiodic lattices. 
For example, according to the concept of $E(r)$, we can now interpret the origin of the energy gaps of localized states. 
In Eq.~\eqref{AAH_real_space_BZ}, when $t=0$, all lattice sites become isolated, densely arranged in the RBZ. In this case,  each site exhibits an on-site energy of $V_r=\cos(G_2\cdot r)$, which corresponds to the band $E_0(r)$, i.e., the pristine $E(r)$. A non-zero $t$ here acts similarly to a quasiperiodic potential in Eq.~\eqref{AAH_k_space_BZ}. So,  we can immediately understand the coupling induced gap by considering the replica bands $E_0(r+ma_0)$ as in IEB.  In fact,   when these replica bands exhibits degeneracy with the pristine $E_0(r)$, a non-zero $t$ opens an energy gap in $E(r)$. This example indicates that, analogous to conventional band theory, all the properties of  localized states in quasiperiodic lattices can be well understood through the LSEB theory.

\emph{Generalized AAH model}---In principle, the reduced-coordinate-based  LSEB theory is applicable to all quasiperiodic lattices. The AAH model is in fact a very special quasiperiodic lattice with a self-dual property, where mobility edges do not exist. Here, we present an example with mobility edges: the generalized AAH (GAAH) model\cite{PhysRevLett.114.146601}.

The primary distinction between the GAAH and the AAH model lies in the different functional form of $V_j$, which breaks the self-dual characteristic and leads to the emergence of mobility edges\cite{PhysRevLett.114.146601}. 
The Hamiltonian of GAAH model is 
\begin{equation}{\label{GAAH_real_space_lattic}}
    H_{\mathrm{GAAH}}=t \sum_{j}( c^{\dagger}_{j}c_{j+1}+\mathrm{H.c.})+\sum_jV_j(\beta,\vartheta)n_j,
\end{equation}
where \begin{equation}{\label{GAAH_potential}}
    V_j(\beta,\vartheta)=2\lambda \frac{cos(G_2\cdot R_j+\vartheta)}{1-\beta cos(G_2 \cdot R_j+\vartheta)}.
\end{equation}
$\lambda$ and
$\vartheta$ are the amplitude and the phase offset. Here, except the quasiperiodic potential $V_j$, all other parameters are the same as the AAH model. The mobility edge $E_{\textrm{ME}}$ here depends on the parameter $\beta$ with $\beta E_{\textrm{ME}}=2\textrm{sgn}(\lambda)(|t|-|\lambda|)$\cite{PhysRevLett.114.146601}.

 The quasiperiodic lattices with mobility edges in fact host a unique hybrid band structure.
Due to the presence of mobility edges, the energy spectrum is divided into several parts, some being extended states and some being localized states. Therefore, according to our theory, the extended states, due to their localization in momentum space, can be described in momentum space using the IEB $E(k)$ with $k \in BZ$, while  the localized states, on the other hand, can well be described by the  LSEB $E(r)$ with $r \in \textrm{RBZ}$.

Numerical results are shown in Fig.~\ref{fig3}\cite{supplemental1}. Here, we set $\beta=-0.5$, so $E_{\textrm{ME}}$ is $-0.4t$ (dashed lines). In Fig.~\ref{fig3} (a), we plot the energy spectrum  in momentum space. As we expected, in momentum space, the extended states below $E_{\textrm{ME}}$ can be well described by IEBs (red solid lines), while the localized states  above $E_{\textrm{ME}}$ appears as completely flat bands with mini gaps. Note that, although the concept of the IEB no longer applies to localized states above the mobility edge, we can still calculate the energy spectrum from the Hamiltonian matrix in momentum space. However, this spectrum now includes lots of redundant eigenstates caused by the replica bands, and due to the localized nature, they exhibits a flat band shape\cite{guo}.
In contrast, with the reduced coordinate in real space, see Fig.~\ref{fig3} (b), the energy spectrum above $E_{\textrm{ME}}$ now can be well described by the LSEB (blue lines). And, the extended states below $E_{\textrm{ME}}$ exhibit flat band shape in real space coordinate. 
Fig.~\ref{fig3} (c) plot the corresponding ARPES of the GAAH model, where the extended states shows a clear energy band structure, while localized states above $E_{\textrm{ME}}$ exhibit a flat band like feature.

\emph{Summary}. Our theory can be summarized in Fig.~3(d). Depending on the choice of real or momentum space, and whether lattice indices or reduced coordinates (BZ) are adopted, the Hamiltonian of a quasiperiodic lattice can exhibit four distinct forms. These forms are interconvertible through basis transformation, spiral mappings, and Aubry-André transformations. The extended states are best described by the IEB concept, which is defined over the BZ in momentum space.  For localized states, the optimal approach is to adopt reduced coordinates (RBZ) in real space, enabling a precise description via the LSEB framework. The energy-band-like concept has been generalized to the localized states in quasiperiodic systems.

\begin{acknowledgments}
    This work was supported by the National Natural Science Foundation of China (Grants No.~12141401 and No.~22273029), the National Key Research and Development Program of China (Grants No.~2022YFA1403501 and No.~2022YFA1402400), China Postdoctoral Science Foundation (Grant No. 2024M750984), and Innovation Program for Quantum Science and Technology (Grant No. 2021ZD0302400). 
\end{acknowledgments}

\bibliography{reference}

\clearpage
\onecolumngrid

\newcommand{\bk}{\bm{k}}
\newcommand{\bq}{\bm{q}}
\newcommand{\btk}{\widetilde{\bm{k}}}
\newcommand{\btq}{\widetilde{\bm{q}}}
\newcommand{\br}{\bm{r}}
\newcommand{\cop}{\hat{c}}
\newcommand{\dop}{\hat{d}}
\newcommand{\xmark}{\ding{55}}
\def\Red#1{\textcolor{red}{#1}}
\def\Blue#1{\textcolor{blue}{#1}}

\begin{center}
\textbf{\large Supplementary Materials for: Theory of Localized States in Quasiperiodic Lattices}
\end{center}

\setcounter{equation}{0}
\setcounter{figure}{0}
\setcounter{table}{0}
\setcounter{page}{1}
\makeatletter
\renewcommand{\theequation}{S\arabic{equation}}
\renewcommand{\thefigure}{S\arabic{figure}}
\renewcommand{\bibnumfmt}[1]{[S#1]}

\section{I. The HAMILTONIAN MATRIX OF THE AAH MODEL IN MOMENTUM SPACE}

As shown in Eq.~\eqref{AAH_k_space_BZ} of the main text, we derive the k-space Hamiltonian of AAH model using the Bloch waves of the atomic chain,  ($\cdot \cdot \cdot, \ket{k-G_2}, \ket{k}, \ket{k+G_2}, \cdot \cdot \cdot$), as the basis. Here, $\ket{k}=\dfrac{1}{N}\sum_j \exp(ik\cdot R_j)\ket{j}$ is the eigenstate of the atomic chain, and $T_k=2t\cos(k\cdot a_0)$ is the kinetic energy of the Bloch wave. When $\alpha$ is an irrational number, the Hamiltonian matrix becomes an infinite dimensional tridiagonal matrix 
\begin{equation}\label{H_AAH}
    H_{AAH}=\begin{bmatrix}
    &  \ddots            &               &            &                   &\\
          &\dfrac{V_0}{2} & T_{k-G_2}    & \dfrac{V_0}{2}             &             &\\
          &              & \dfrac{V_0}{2} & T_{k}   & \dfrac{V_0}{2}                &\\
          &              &               &\dfrac{V_0}{2}  & T_{k+G_2}&  \dfrac{V_0}{2}             \\
          &              &               &              &                 &\ddots
\end{bmatrix}.
\end{equation}

\section{II.THE LIMITATION OF IEB THEORY FOR DESCRIBING THE LOCALIZED STATES}
\begin{figure}[h!]
    \centering
    \includegraphics[scale=1.5]{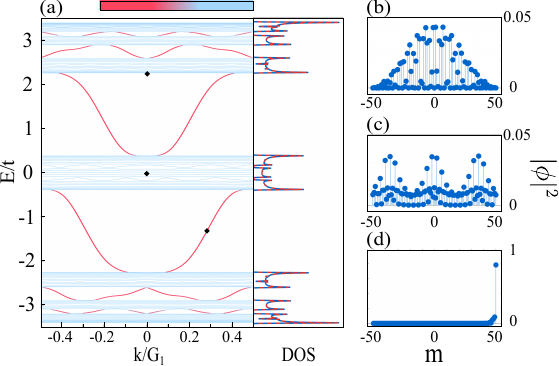}
    \caption{The spectrum and wave function distribution of AAH model with $V_0=3t$. (a) is the calculated spectrum as a function of $k$, where we use color to denote the $\mathrm{log}_{10}(\mathrm{IPRM})$ of each eigenstate, with $n_c=50$, $n_e=5$. The right panel provides the calculated DOS, where red lines are obtained from the spectrum calculated with a larger truncation $n_c=250$, $n_e=2$, and blue lines are got by directly diagonalizing a large enough lattice in real space. 
              (b-d) show the wave function distributions in k-space for the three eigenstates corresponding to the three black dots in (a), arranged sequentially from top to bottom. $m$ is the index of the Bloch basis $|k+mG_2\rangle$.
           }
    \label{figs2}
\end{figure}
The IEB theory can give a well description about the extended states in quasiperiodic lattices. However, we can not define the concept of IEB for the localized states in quasiperiodic lattice. It is because that, to define IEB, we assume that the eigenstates are localized in momentum space, a basic feature of the extended states.  This assumption becomes invalid for the localized states in quasiperiodic lattices. 

The limitation of the IEB theory for describing the localized states are shown clearly in Fig.~\ref{figs2}.  With the Hamiltonian matrix in Eq.~(2) of the main text, we can still calculate the spectrum of the eigenstates as a function of $k$  when $V_0 \geq 2t$, see  Fig.~\ref{figs2} (a). 
We further illustrate the localization feature of these eigenstates in Fig.~\ref{figs2} (a) by plotting their IPRM. The inverse participation ratio (IPR) is a standard measure of localization properties for wave functions in quasiperiodic lattices, and IPRM, i.e. IPR in momentum space,  is the corresponding property calculated in momentum space~\cite{guo}. 
As shown in Fig.~\ref{figs2} (a), the IPRM of eigenstates is color-coded (see color bar), with blue representing states extended in momentum space (localized in real space) and red denoting states localized in momentum space (extended in real space).  
Obviously,  nearly all the eigenstates become localized (blue lines),  whose wave functions are then extended in momentum space, see Fig.~\ref{figs2} (b,c). Therefore, the IEB can not be identified in this case. Note that, due to the localized feature, the calculated spectrum exhibit flat band shape. Fortunately, though the IEB concept is no longer applicable,  we can still use the calculated spectrum to get the correct DOS with the help of proper weight factor of each eigenstates, see Fig.~\ref{figs2} (a). The weight factors of eigen functions are in fact determined by  the symmetry of the quasiperiodic lattice Hamiltonian, and this applies to both localized and extended states~\cite{guo}.   
Note that, due to the truncation of the Hamiltonian matrix,  some edge states in momentum space exist, see red lines Fig.~\ref{figs2} (a), and the corresponding wave function is given in Fig.~\ref{figs2} (d). 
The formula of the IPRM is given in next section.

\section{III. THE INCOMMENSURATE ARPES THEORY for QUASIPERIODIC LATTICE }
The ARPES theory for incommensurate systems, like the AAH model, has been developed in our recent work\cite{guo}. Here, we give a short introduction to the ARPES calculation method.

For a quasiperiodic system, the photoemission current $I$ measured in an ARPES experiment can be expressed as
\begin{equation}
    I(\bm{p},E) \propto \dfrac{1}{N_E}\sum_{\bm{k} \in BZ} \sum_i |\bra{\bm{p}} H_{int} \ket{\Phi^{(i)}(\bm{k})}|^2 \delta(E-\epsilon^{(i)}(\bm{k})),
\end{equation}
where $i$ refers to the band index, and $1/N_E$ is the weighting factor for each eigenstate $\ket{\Phi^{(i)}(\bm{k})}$, which depends on the truncation of the Hamiltonian matrix. Specifically, $N_E=2(n_c-n_e)+1$, $n_c$ and $n_e$ are the truncation value and boundary width, respectively\cite{he2024energy}. Note that the above formula is applicable regardless of whether the eigenstates are extended or localized.

Then, we consider the non-interacting case, i.e. sudden approximation, and take the simplest approximation, assuming a free electron final state $\ket{\bm{p}}$\cite{MOSER201729,sobota2021angle}. $\ket{\Phi^{(i)}(\bm{k})}$ is  calculated by diagonalizing the Hamiltonian matrix,
\begin{equation}
    \ket{\Phi^{(i)}(\bm{k})}=\sum_{m\in \mathbb{Z}} \phi^{(i)}_m(\bm{k})\ket{\bm{k}+m\bm{G_2}},
\end{equation}
where $\phi^{(i)}_m(k)$ is the coefficient of the basis $\ket{\bm{k}+m\bm{G_2}}$. With the dipole approximation\cite{sobota2021angle}, the light-matter interaction Hamiltonian can be described as $\mathrm{H}_{int}$:
\begin{equation}
    \mathrm{H}_{int} \sim \dfrac{e}{mc} \bm{A}\cdot \hat{\bm{p}}=-\dfrac{ie\hbar}{mc} \bm{A} \cdot \nabla,
\end{equation}
with $\hat{\bm{p}}$ is the photoelectron momentum operator, $\bm{A}$ is the vector potential of the incoming photon, which is assumed to be constant in space. Consequently,
\begin{align}
    \begin{aligned}
        I(\bm{p},E) &\propto \dfrac{1}{N_E}\sum_{\bm{k}\in BZ} \sum_i |\bra{\bm{p}} H_{int} \ket{\Phi^{(i)}(\bm{k})}|^2 \delta(E-\epsilon^{(i)}(\bm{k})) \\
        &\propto \dfrac{1}{N_E} \sum_{\bm{k}\in BZ} \sum_i  |\bra{\bm{p}} \bm{A}\cdot \nabla \ket{\Phi^{(i)}(\bm{k})}|^2 \delta(E-\epsilon^{(i)}(\bm{k}))  \\
        &=\dfrac{1}{N_E} \sum_{\bm{k}\in BZ} \sum_i  |\bm{A}\bra{\bm{p}} \nabla \ket{\Phi^{(i)}(\bm{k})}|^2 \delta(E-\epsilon^{(i)}(\bm{k}))  \\
        &= \dfrac{1}{N_E} |\bm{A}\cdot \bm{p}|^2 \sum_{\bm{k}\in BZ} \sum_i  |\langle \bm{p} |\Phi^{(i)}(\bm{k})\rangle|^2 \delta(E-\epsilon^{(i)}(\bm{k})),
    \end{aligned}
\end{align}
where  $\nabla^\dag=-\nabla$ is used. The inner product can be calculated as
\begin{equation}
    \langle \bm{p} |\Phi^{(i)}(\bm{k})\rangle=\sum_m \phi^{(i)}_m(\bm{k}) \langle \bm{p}|\bm{k}+m\bm{G_2}\rangle.
\end{equation}
Note that the Bloch basis can be written as combination of atomic orbitals, $|k\rangle= \dfrac{1}{N}=\sum_j \exp(i\bm{k}\cdot \bm{R_j})|j\rangle$, where $|j\rangle$ is the atomic orbital of the $\bm{R_j}$ site. Then we have
\begin{align}
    \begin{aligned}
        \langle \bm{p}|\bm{k}+m\bm{G_2}\rangle
        &=\sum_j e^{i(\bm{k}+m\bm{G_2})\cdot \bm{R_j}}\langle \bm{p}|j\rangle  \\
        &=\sum_j e^{i(\bm{k}+m\bm{G_2})\cdot \bm{R_j}} \int d\bm{r} e^{-i\bm{p}\cdot\bm{r}} \psi(\bm{r}-\bm{R_j})  \\
        &=\sum_j e^{i(\bm{k}+m\bm{G_2}-\bm{p})\cdot \bm{R_j}} \int d\bm{r} e^{-i\bm{p}\cdot(\bm{r}-\bm{R_j})} \psi(\bm{r}-\bm{R_j})  \\
        &= \psi(\bm{p}) \delta_{\bm{p}_{||},\bm{k}+m\bm{G}_2+n\bm{G}_1},
    \end{aligned}
\end{align}
where $\psi(\bm{p})=\int d\bm{r} e^{-i\bm{p}\cdot\bm{r}} \psi(\bm{r})$ and $\psi(\bm{r}-\bm{R_j})$ is the wave function of the atomic orbital centered at $\bm{R}_j$. Finally, we get the ARPES formula
\begin{equation}{\label{arpes}}
    I(\bm{p},E)  \propto \dfrac{1}{N_E} |\bm{A}\cdot \bm{p}|^2 |\psi(\bm{p})|^2 \sum_{\bm{k} \in BZ} \sum_i \delta(E-\epsilon^{(i)}(\bm{k})) \times |\sum_m \phi_m^{(i)}(\bm{k}) \delta_{\bm{p_{||}},\bm{k}+m\bm{G_2}+n \bm{G_1}}|^2.
\end{equation}
Here, $\bm{p}$ is the photoelectron momentum, $\epsilon^{(i)}(\bm{k})$ are the corresponding eigenvalues.  $\psi(\bm{p})=\int d\bm{r} \exp(-i\bm{p}\cdot\bm{r}) \psi(\bm{r})$ is the Fourier transform of the atomic orbital on lattice. $|\bm{A}\cdot \bm{p}|^2$ depends on the polarization vector $\bm{A}$ of the incoming photon. 

Finally, we give the formula of the IPRM, i.e., the inverse participation ratio in momentum space,  which is used to reflect the localized feature of the wave function in momentum space.
\begin{equation}{\label{IPRM}}
    \mathrm{IPRM}(|\Phi^{(i)}(\bm{k})\rangle)=\sum_m |\phi_m^{(i)}(\bm{k})|^4
\end{equation}

\section{IV. Basis transformation Formula for the Hamiltonian}
As mentioned in the main text, the Hamiltonian of the AAH model have different forms, which  can be transformed into each other through a basis transformation together with spiral mapping. 
For example, Eq.~(3) is the k-space Hamiltonian of the AAH model, which is represented as a coupled chain in momentum space, and each  site  corresponds to a Bloch wave $|K_m\rangle$ with $m$ as a site index.
Meanwhile, Eq.~(4) is the real space Hamiltonian of the AAH model, where each site represents a wannier function $|r\rangle$ (atomic orbital), indicating by the reduced coordinate $r=\widetilde{R}_j$ with $\widetilde{R}_j=R_j$ mod $b_0$ and $R_j=j\cdot a_0$. 

We can transform the Hamiltonian of AAH model from the form in Eq.~(3) to the form in Eq.~(4) through a change of basis, 
\begin{equation}\label{basistransform2}
  |r\rangle = \frac{1}{\sqrt{N}} \sum_{m} e^{-i K_m \cdot r} |K_m\rangle.  
\end{equation}
Here, we provide a brief interpretation about the above formula. Let us start with the well-known relation,
\begin{equation}\label{basistransform1}
    |j\rangle=\frac{1}{\sqrt{N}}\sum_{k \in \textrm{BZ}} e^{-ik\cdot R_j} |k\rangle,
\end{equation}
 where $|j\rangle$ is  the atomic orbital function and and $R_j$ is position of the j-th atom, and $\ket{k}$ is the corresponding Bloch wave. First, according to the spiral mapping relation, the coupled Bloch waves have two identical index, i.e., $\ket{k=\widetilde{K}_m}$, where $K_m=mG_2$ and $\widetilde{K}_m=K_m$ mod $G_1$. So, due to this one-to-one correspondence, we have $\ket{k}=\ket{K_m}$. Then, for the atomic orbital $\ket{j}$, we also have two identical index. With similar reason, we also have $\ket{j}=\ket{r}$. Note that, for the reduced coordinate, $r=R_j+p\cdot b_0$, where $p$ is an integer to ensure $r$ is reduced to the RBZ.  Meanwhile, $k=K_m+nG_1$, where $n$ is an integer to make $k$ to be confined in BZ. With these relations,  we can directly prove that,
 \begin{equation}
    e^{-ik\cdot R_j}=e^{-iK_m\cdot r}.
 \end{equation}
 At last, due to the spiral mapping relation in momentum space, we note that the summation over the BZ is equivalent to the summation over the whole coupled momentum chain. 
Finally, based on all the relations above, Eq.~\eqref{basistransform1} transforms to the form of Eq.~\eqref{basistransform2}. This is just the basis transformation formula used in  the main text.

\section{V.Calculation Details for the Localized States in the AAH Model}

\subsection{A. How to calculate the localized states of AAH model}
To calculate the localized states in AAH model, the key idea is based on the duality relation between the momentum and real spaces. 

Through Eq.~(3) and Eq.~(4) of  the main text, we observe a clear duality relationship between the Hamiltonian of the AAH model in momentum space and real space. Within the IEB theory, we use the momentum-space Hamiltonian, i.e., Eq.~(3) in the main text, and apply the IEB method to solve the energy bands of extended states (extended in real space and localized in momentum space). Eq.~(4), on the other hand, represents the real-space form of the AAH model Hamiltonian and exhibits a perfect duality with Eq.~(3): 
\begin{enumerate}
    \item Eq.~(3) is based on Bloch waves $\ket{k}$, where $k$ is the crystal momentum defined in the BZ, while Eq.~(4) is based on Wannier waves $\ket{r}$, where $r$ is the reduced coordinate defined in the RBZ;
    \item Eq.~(3) and Eq.~(4) share the same Hamiltonian matrix form. The Hamiltonian matrix form of Eq.~(3) is given in Eq.~\eqref{H_AAH}, while the Hamiltonian matrix of Eq.~(4) is
\begin{equation}\label{AAH_LSEB_Matrix}
    H_{AAH}=\begin{bmatrix}
    &  \ddots            &               &            &                   &\\
          &t & V_{r-a_0}    & t             &             &\\
          &              & t & V_{r}   & t                &\\
          &              &               &t  & V_{r+a_0}&  t             \\
          &              &               &              &                 &\ddots
\end{bmatrix}
\end{equation}
where the Wannier functions  $\{\cdots,\ket{r-a_0},\ket{r},\ket{r+a_0},\cdots \}$ are used as basis, and $V_r=V_0 \cos(G_2\cdot r)$. Note that the correspondence relations between the two matrix:  $t \sim \frac{V_0}{2}$, $V_r \sim T_k$, $a_0 \sim G_2$, $r \sim k$. 
    
    \item Extended states, calculated with IEB theory,  are essentially  \textit{the localized states in momentum space}. So, it is not surprising that \textit{the localized states in real space} can be calculated with similar process based on Eq.~(4).  
\end{enumerate}

Given this duality, we can directly map Eq.~(4)  onto Eq.~(3). By following a procedure analogous to the IEB-based treatment of Eq.~(3), we can similarly solve for the localized states defined on the reduced coordinate $r$. The resulting \textit{dispersion relationship} $E(r)$ then corresponds precisely to what we define as the \textit{localized state energy band}.

\subsection{B. The DOS formula}
In the main text, we provide the formula to calculate the DOS of localized states in the AAH model, see Eq.~(5). Here, we give a short explain about this formula.

As indicated in the main text,  each localized state in AAH model actually predominantly localizes around one lattice site, which implies we can use the lattice site to label the localized state. For example, the LSEB $E(r)$ in fact means the  energy of the localized state around site $r=\widetilde{R}_j$ (j-$th$ site). So, the DOS can be calculated via accounting all the localized states in the atomic chain
\begin{equation}
    \rho(\epsilon)=\lim_{L\rightarrow \infty}\dfrac{1}{L}\sum_{j=0}^{L-1}\delta(\epsilon-E(\widetilde{R}_j)),
\end{equation}
where $L$ is the total number of the sites of the chain.  Of course, it is hard to directly calculate DOS with this formula, since the atomic chain is infinitely long. 
We further notice that the spiral mapping relation  maps all the lattice sites onto a compact circle, i.e., the RBZ. Importantly, when $\alpha=a_0/b_0$ is irrational, the spiral mapping process of all the lattice sites $\{ \widetilde{R}_j \}$ is in fact equivalent to the celebrated irrational rotation problem\cite{irrational_rotation}. Then, by Birkhoff's ergodic theorem\cite{birkhoff,walters2000introduction}, we get the final formula for the DOS calculation, 
\begin{equation}
    \rho(\epsilon)=\lim_{L\rightarrow \infty}\dfrac{1}{L}\sum_{j=0}^{L-1}\delta(\epsilon-E(\widetilde{R}_j))=\dfrac{1}{b_0}\int_0^{b_0}dr \delta(\epsilon-E(r)).
\end{equation}
Now the DOS of the localized states can be calculated via a simple numerical integration.

\section{VI.THE calculation DETAILs of the  GAAH MODEL}
Here we provide some  calculation details of the GAAH model. 

First, we derive the k-space Hamiltonian. Note that the incommensurate periodic potential can be expressed through a Fourier series expansion
\begin{equation}{\label{plain_wave_expansion}}
    V_j=\sum_{G} V_{G} \cdot e^{iG \cdot R_j},
\end{equation}
with $G=l\cdot G_2$, $l\in \mathbb{Z}$. 
Then, with the Bloch waves $|k\rangle = \frac{1}{\sqrt{N}} \sum_j \exp(-ik\cdot R_j) |j\rangle$ as basis, we have
\begin{equation}
    H_{\mathrm{GAAH}}=\sum_{k\in \left[0,G_1 \right)}[T_k+V_\beta]n_k+ \sum_{k\in \left[0,G_1 \right)} \sum_{G\neq 0} V_G(c^\dag_k c_{k+G}+ \mathrm{H.c.}).
\end{equation}
Here, $T_k=2t\cos(k\cdot a_0)$, and we define $V_\beta \equiv V_{G=0}$, which only depends on  the value of $\beta$. $c_k$ is the electron annihilation operator of the Bloch state $\ket{k}$. $V_G$ is the expansion coefficients  
\begin{equation}{\label{expansion_coeffient}}
    V_G=\dfrac{1}{b_0}\int_0^{b_0} dr e^{-iG \cdot R_j}V_{j}(\beta).
\end{equation}
Finally, we can get the the Hamiltonian matrix of GAAH model 
\begin{equation}\label{k_space_Matrix_GAAH}
    H_{GAAH} = \left[
    \begin{array}{ccccccccccc}
        \ddots &        &        &   \vdots &         &         &          &          &         &         &    \\
               & \ddots &        &  V_{2G_2}& \vdots  &         &          &          &         &         &    \\   
               &        & \ddots &  V_{G_2} & V_{2G_2}& \vdots  &          &          &         &         &    \\
               &        &        &T_{k-2G_2}& V_{G_2} & V_{2G_2}& \vdots   &          &         &         &    \\
               &        &        &V_{-G_2}  &T_{k-G_2}& V_{G_2} & V_{2G_2} &  \vdots  &         &         &    \\
               &        &        & V_{-2G_2}& V_{-G_2}& T_{k}   & V_{G_2}  &  V_{2G_2}&         &         &    \\
               &        &        &  \vdots  &V_{-2G_2}&V_{-G_2} & T_{k+G_2}&  V_{G_2} &         &         &    \\
               &        &        &          &\vdots   &V_{-2G_2}& V_{-G_2} &T_{k+2G_2}&         &         &    \\
               &        &        &          &         &\vdots   & V_{-2G_2}& V_{-G_2} &\ddots   &         &    \\
               &        &        &          &         &         & \vdots   & V_{-2G_2}&         &\ddots   &    \\
               &        &        &          &         &         &          & \vdots   &         &         &\ddots
    \end{array}
    \right] + V_\beta
\end{equation}
Note that, due to the existence of mobility edge, localized and extended states always coexist in the GAAH model. 
Therefore, although we can always obtain all its eigenstates (as functions of k) of this momentum space Hamiltonian matrix, the IEBs can be identified only for extended states. The energy spectrum of localized states exhibit a flat band shape. And we can not define IEB for these localized states, since their wave functions are extended in momentum space.

Similar as the AAH model discussed in the main text, the k-space Hamiltonian can also be expressed as a momentum space lattice via the spiral mapping relation.  Note that all the coupled Bloch waves $\ket{k}$ are  on a coupled chain in momentum space, i.e., $k=K_m\equiv mG_2$.  So, k-space the Hamiltonian can be rewrited as  
\begin{equation}{\label{GAAH_k_space_lattice_H}}
    H_{\mathrm{GAAH}}=\sum_{m}[T_m+V_\beta]n_m+ \sum_m \sum_{l \neq 0} V_G(c^\dag_m c_{m+l}+ \mathrm{H.c.}).
\end{equation}
 Here $c_m \equiv c_{K_m}$ is the annihilation operator of Bloch state $\ket{K_m}$, and $T_m = 2t\cos(K_m \cdot a_0)$. We note that such Hamiltonian in fact imply that the GAAH model does not have the self-duality feature.
 
 Now, we derive the real space Hamiltonian expressed by the reduced coordinate. Following the approach used for the AAH model, we still take the basis transformation $\ket{r} = \frac{1}{\sqrt{N}}\sum_m e^{iK_m\cdot r} \ket{K_m}$. Substituting this into Eq.~\eqref{GAAH_k_space_lattice_H} yields
\begin{equation}{\label{GAAH_real_space_BZ_H}}
    H_{\mathrm{GAAH}}=\sum_{r\in \left[0,b_0 \right)} t(c_r^\dag c_{r+a_0}+h.c.)+\sum_{r\in \left[0,b_0 \right)} V_r(\beta)n_r.
\end{equation}
Here $r$ is the reduced coordinate defined on RBZ and $V_r(\beta) = 2\lambda \cos(G_2 \cdot r)/(1 - \beta\cos(G_2 \cdot r))$, $c_r \equiv c_{R_j}$ is the annihilation operators of atomic orbital functions.
The matrix form of Eq.~\eqref{GAAH_real_space_BZ_H} is
\begin{equation}\label{real_space_Matrix_GAAH}
    H_{GAAH}=\begin{bmatrix}
 & \ddots &         &                   &                &            &                      &       & &\\   
 &  &  \ddots   &    t        &                  &           &                      &      & &\\
 &  &       &  V_{r-2a_0}     &     t       &             &                   &   &      &\\
  &  &      &t & V_{r-a_0}    & t             &              &   &         &\\
 &   &      &               & t & V_{r}   & t               &    &      &\\
  &   &     &              &                &t  & V_{r+a_0}     &  t    &       &\\
 &    &     &              &               &               &    t  &V_{r+2a_0} &      &\\
  &   &     &               &              &              &              &   t &\ddots  &\\
  &   &     &              &               &               &            &            &       &\ddots
\end{bmatrix}.
\end{equation}
Then, with the real space Hamiltonian above, we can calculate the LSEB of the GAAH model using the same scheme proposed in the main text.


\section{VII.THE DUAL TRANSFORM of AAH MODEL}

Here, we provide a dual transform formula
\begin{equation}
    |\bm{k} \rangle=\frac{1}{\sqrt{N}}\sum_{\bm{r}}\exp(i \bm{k} \cdot \bm{r} )|\bm{r}\rangle,
\end{equation}
which is a new dual transform between the two forms of the Hamiltonian of AAH model, i.e., Eq.~(2) and Eq.~(4).

\end{document}